\documentclass[twocolumn,secnumarabic, showpacs, superscriptaddress,amssymb,english,aps, prb,floatfix, longbibliography]{revtex4-1}

\usepackage{graphicx}
\usepackage{SIunits}
\usepackage{epstopdf}

\begin{document}

\title{Spin-dependent direct gap emission in tensile-strained Ge films on Si substrates}%

\author{E. Vitiello}%
\email{e.vitiello1@campus.unimib.it}
\affiliation{LNESS and Dipartimento di Scienza dei Materiali, Universit\`a degli Studi di Milano-Bicocca, via Cozzi 55, I-20125 Milano, Italy}
\author{M. Virgilio}
\affiliation{Dipartimento di Fisica, Universit\`a di Pisa, Largo Pontecorvo 3, I-56127 Pisa, Italy}
\author{A. Giorgioni}
\affiliation{LNESS and Dipartimento di Scienza dei Materiali, Universit\`a degli Studi di Milano-Bicocca, via Cozzi 55, I-20125 Milano, Italy}
\author{J. Frigerio}
\affiliation{LNESS and Dipartimento di Fisica del Politecnico di Milano, Polo di Como, via Anzani 42, I-22100 Como, Italy}
\author{E. Gatti}
\author{S. De Cesari}
\author{E. Bonera}
\author{E. Grilli}
\affiliation{LNESS and Dipartimento di Scienza dei Materiali, Universit\`a degli Studi di Milano-Bicocca, via Cozzi 55,
I-20125 Milano, Italy}
\author{G. Isella}
\affiliation{LNESS and Dipartimento di Fisica del Politecnico di Milano, Polo di Como, via Anzani 42, I-22100 Como, Italy}
\author{F. Pezzoli}
\affiliation{LNESS and Dipartimento di Scienza dei Materiali, Universit\`a degli Studi di Milano-Bicocca, via Cozzi 55,
I-20125 Milano, Italy}
\date{\today}%

\begin{abstract}
The circular polarization of direct gap emission of Ge is studied in optically-excited tensile-strained Ge-on-Si heterostructures as a function of doping and temperature. Owing to the spin-dependent optical selection rules, the radiative recombinations involving strain-split light ($\mathrm{c\Gamma-LH}$) and heavy hole ($\mathrm{c\Gamma-HH}$) bands are  unambiguously resolved. The fundamental $\mathrm{c\Gamma-LH}$ transition is found to have a low temperature circular polarization degree of about 85\% despite an off-resonance excitation of more than 300 meV. By photoluminescence (PL) measurements and tight binding calculations we show that this exceptionally high value is due to the peculiar energy dependence of the optically-induced electron spin population. Finally, our observation of the direct gap doublet clarifies that the light hole contribution, previously considered to be negligible, can dominate the room temperature PL even at low tensile strain values of $\approx 0.2\%$.
\end{abstract}

\pacs{72.25.Fe, 72.25.Rb, 78.55.Ap, 78.20.-e}

\maketitle

Strain offers an effective degree of freedom for tailoring the band structure of semiconducting materials, opening up interesting physical phenomena and advanced application perspectives. \cite{Kato04,Pereira09,Bedell14} In the context of spin physics, strain can be exploited to modify relaxation mechanisms and thus spin lifetimes.\cite{dyakonov, zutic:review2004, dery:PRB2013} Furthermore, the zone-center removal of light (LH) and heavy (HH) hole valence band (VB) degeneracy can significantly improve the optical generation of non-equilibrium spin populations. \cite{zutic:review2004} Transfer of photon angular momentum to a material near the limit of complete spin polarization has been demonstrated in strained III-V semiconductors.\cite{zutic:review2004,Tessler:APL94} This has been achieved through resonant absorption of circularly polarized light at an excitation energy that matches the fundamental strain-split transition.\cite{zutic:review2004,Tessler:APL94} Similar possibilities have been pointed out in indirect gap systems based on group IV materials, such as Ge quantum wells under compressive strain.\cite{Virgilio09, pezzoliPRL2012, lange:PRBR2012, giorgioni:APL2013} 
Ge offers indeed an intriguing spin physics \cite{dery:PRB2013, loren:PRB2011, Morrison14}, thanks to its multi-valley conduction band (CB) \cite{pezzoliPRB2013}, the relatively large spin-orbit coupling \cite{zutic:review2004} and long electron spin-lifetime.\cite{giorgioni:APL2014} 

\begin{figure*} [t]
\includegraphics[width=2\columnwidth]{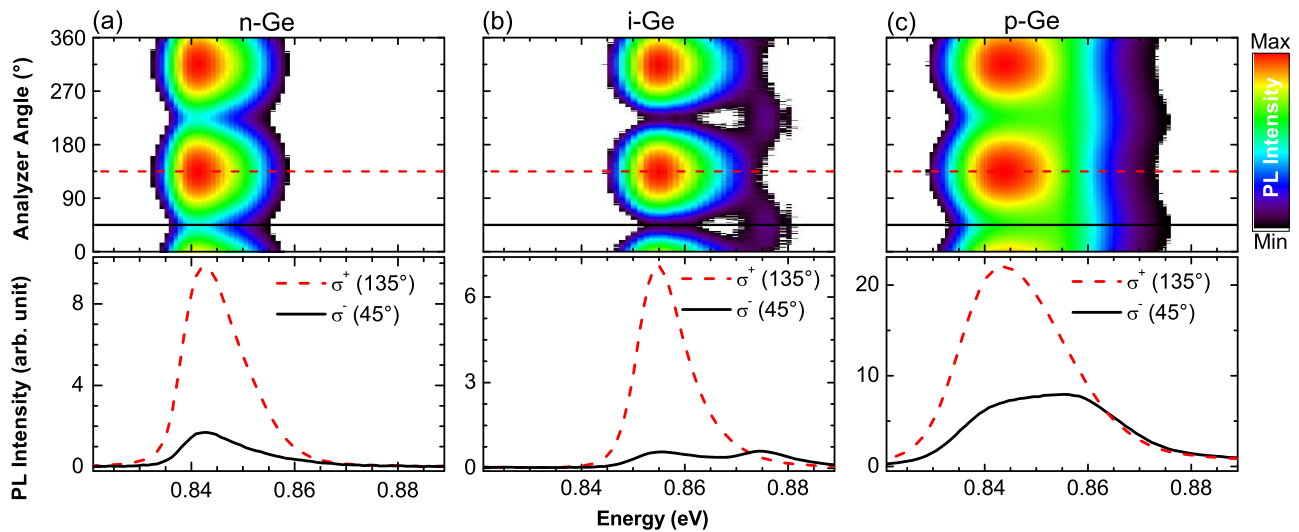}
\caption{\label{fig1}Upper panels: Color-coded intensity maps of the photoluminescence (PL) measured as a function of the analyzer angle at 4 K in tensile-strained n-Ge (a), i-Ge (b), p-Ge (c). \cite{pezzoliPRB2013,goldstein} Lower panels: 4 K PL spectra of n-Ge (a), i-Ge (b) and p-Ge (c) corresponding to $\sigma^{-}$ (analyzer angle at $45^{o}$, black solid line) and $\sigma^{+}$ ($135^{o}$, red dashed line) polarization components.}
\end{figure*}

In the last few years, it has been recognized that heteroepitaxy of thin films of Ge on Si substrates (Ge-on-Si) spontaneously introduces a biaxial tensile strain, $\varepsilon_\parallel$, in the epitaxial layer upon cooling from the high growth temperature. \cite{ishikawa:JAP2005, liu:OL2010,camacho:OE2012} Such strain, although relatively small, i.e. $\varepsilon_\parallel \approx 0.2$\%, is sufficient to boost the direct gap emission as exemplified by the demonstration of lasing in Ge-on-Si.\cite{liu:OL2010,camacho:OE2012} More recently, the prominence of this strained heterosystem has been further strengthened by transport measurements, which showed room temperature spin diffusion lengths in heavily doped samples up to 660 nm.\cite{Dushenko15} Despite such advances, several questions remain unanswered, such as the impact of non radiative recombination on carrier dynamics and optical gain. \cite{Virgilio13} In particular, no systematic optical study has been focused on the spin properties, although the understanding of carrier and spin kinetics in strained Ge-on-Si is of both fundamental interest and central importance for applications in spintronics and photonics.

In this work we study the effect of tensile strain on the spin-dependent radiative recombination of Ge by measuring, upon optical orientation of the carrier spin, \cite{Rioux10} the polarization of the direct gap PL. This allows us to spectrally resolve emission involving light (c$\Gamma$-LH) and heavy holes (c$\Gamma$-HH) over a wide temperature and doping range. Surprisingly, for an excess pump energy of about 300 meV with respect to the fundamental c$\Gamma$-LH transition, we measure a low temperature PL polarization degree ($\rho$) as high as 85\%. This is the highest value reported in bulk Ge films. \cite{Bottegoni11, Ferrari13, Pezzoli15} Relying on tight binding calculations, we demonstrate that this polarization degree is due to the non-trivial energy dependance predicted for the non-equilibrium electron spin population. Finally, our results elucidate the so-far-overlooked role of the occupancy of strain-split LH and HH hole states in determining the direct gap PL spectrum. \cite{elkurdi:APL2010, boztug:NN2014, harris:JAP2014} 

We studied a set of Ge films grown on (001)Si substrates by low energy plasma enhanced chemical vapor deposition. The 1.5$-$2 $\mu \mathrm{m}$ thick Ge layers were deposited at 500$\mathrm{^o}$C at a rate of 4.2 nm/s and in-situ doped as follows: One Ge:P and one Ge:B sample, dubbed n-Ge and p-Ge respectively, with the same doping content of $4.5 \times 10^{17}$ $\mathrm{cm^{-3}}$, and one not-intentionally doped sample (i-Ge) that has an estimated impurity concentration in the low $10^{15}$ $\mathrm{cm^{-3}}$ range, due to the p-type background doping of the reactor. To ensure a tensile strain in the epilayers, all the Ge-on-Si heterostructures were subject at the end of the growth to an in-situ 6-fold cyclic annealing between 600$\mathrm{^o}$C and 780$\mathrm{^o}$C. We deposited and annealed under the same conditions described above a set of four additional samples with a B concentration increasing as follows: $3\times 10^{16}$ $\mathrm{cm^{-3}}$, $2\times 10^{17}$ $\mathrm{cm^{-3}}$, $5\times 10^{17}$ $\mathrm{cm^{-3}}$ and $9\times 10^{17}$ $\mathrm{cm^{-3}}$. The doping level of the samples has been estimated by room temperature resistivity measurement. We verified, by means of Raman spectroscopy carried out at room temperature, that all the samples have the same biaxial tensile strain value of 0.2\%.\cite{pezzoli:ramanstrain}

\begin{figure*} 
\centering
\includegraphics[width=2\columnwidth]{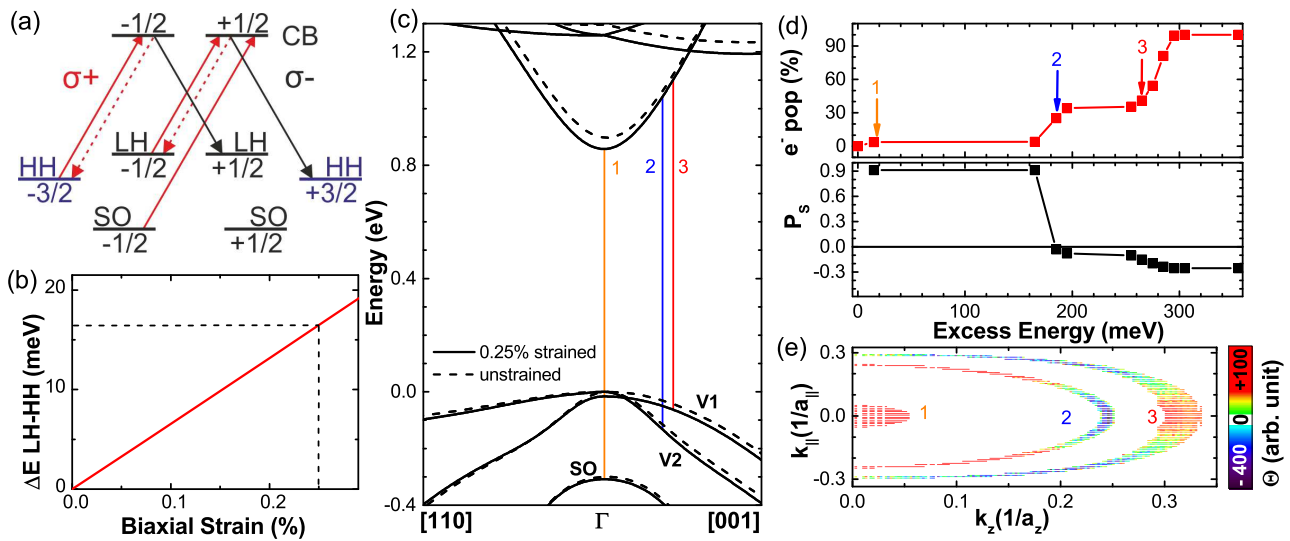}
\caption{\label{fig2}(a) Scheme of the dipole-allowed transitions for right- ($\sigma^{+}$, red lines) and left-handed ($\sigma^{-}$, black lines) circularly polarized light involving $\Gamma$-point heavy hole (HH), light hole (LH), split-off (SO) valence (VB) and conduction (CB) band levels.\cite{zutic:review2004} Fractional values represent the quantum numbers of the component of the angular momentum, $\mathrm{J_z}$, along the direction of light propagation, z. (b) Strain-induced LH-HH splitting calculated via tight binding method at $\Gamma$ as a function of a biaxial tensile strain. (c) Low temperature band structure of relaxed bulk Ge (dashed lines) and of Ge with 0.25\% biaxial tensile strain (solid lines). The latter mimics the Ge-on-Si heterostructure. Vertical transitions from 1 to 3 are promoted respectively by the absorption of photons at $\hbar \mathrm{\omega_{p}}$ = 1.165 eV from the topmost (V1) and second (V2) out-of-zone-center VB states and from the SO states near $\Gamma$. (d) Relative population (upper panel) and spin polarization $\mathrm{P_s}$ (bottom panel) as a function of the excess energy of CB electrons optically promoted by $\sigma^{+}$ excitation. (e) Color-coded map of the photogenerated electronic spin population $\Theta$ as function of the components of the electron momentum.}
\end{figure*}

Polarization-resolved PL measurements were carried out in a back-scattering geometry. The samples were excited by the $\hbar \mathrm{\omega_{p}}$ = 1.165 eV line of a cw right-handed circularly polarized ($\sigma^{+}$) Nd:YVO$_{4}$ laser. The excitation power density was about $4 \mathrm{kW/cm^{2}}$. The PL was probed by the polarization-sensitive InGaAs-based detection system described in detail in Refs. \onlinecite{pezzoliPRL2012} and \onlinecite{pezzoliPRB2013}. We recall here that the amplitude modulation of the PL peaks as a function of the relative orientation (analyzer angle) between an optical retarder and a polarizer provides a full characterization of the degree and state of light polarization.\cite{pezzoliPRB2013,goldstein} 

Figure~\ref{fig1} reports the 4 K PL ascribed to the radiative recombination between electrons photogenerated at the $\Gamma$-valley of the CB and holes at the top of the VB of Ge. The upper panels of Fig.~\ref{fig1} show the color-coded maps of the PL intensity as a function of the analyzer angle for n-Ge [Fig.~\ref{fig1} (a)], i-Ge [Fig.~\ref{fig1}(b)] and p-Ge [Fig.~\ref{fig1}(c)]. It is worth noting that the intensity oscillations occur in phase among all the samples and have a period equal to $\pi$, which stems for circularly polarized PL having the same well-marked helicity.\cite{pezzoliPRL2012, pezzoliPRB2013, goldstein} The lower panels of Fig.~1 better emphasize such result by reporting line scans from the maps at the analyzer angles of $45^{o}$ (black solid curve) and $135^{o}$ (red dashed curves). The latter (former) corresponds to $\sigma^{+}$ ($\sigma^{-}$) circularly polarized PL, i.e. emission copolarized (counter polarized) with respect to the excitation. 

In n-Ge a well-defined PL peak can be observed with maximum at 0.843 eV, see Fig.~\ref{fig1} (a). Its energy position is consistent with the shrinking of the band gap with respect to bulk, unstrained Ge caused by the concomitant contributions of tensile strain \cite{ishikawa:JAP2005} and doping.\cite{jain:bgn1991} The $\sigma^{+}$ component of the emission is exceedingly brighter than the $\sigma^{-}$ counterpart, yielding a large PL circular polarization degree\cite{pezzoliPRB2013} $\rho = 70 \pm 1$\%. The reduced doping concentration of i-Ge  blue-shifts the PL peak to 0.855 eV  [Fig.~\ref{fig1}(b)]. The analysis of i-Ge reveals a striking $\rho$ = 85 $\pm 1\%$, hence larger than the one of n-Ge. Unexpectedly, the measurement of the $\sigma^{-}$ component, shown in Fig.~\ref{fig1}(b), unveils the presence of an additional weaker PL line at 0.874 eV. The two peaks observed in i-Ge have indeed opposite circular polarization, with the lowest energy component of the doublet being copolarized with the laser excitation, whilst the highest energy one is counter polarized. Yet the polarization analysis allows us to distinguish two contributions at 0.843 eV and at about 0.86 eV also in Fig.~\ref{fig1}(c), which shows the PL of p-Ge. In this sample the higher energy emission has a larger spectral weight than in the less doped i-Ge, yielding the broadest PL line among those summarized in Fig.~1.

The aforementioned cross-polarized PL features can be ascribed to direct radiative recombinations of electrons with strain-split LH and HH and because of the tensile strain in the Ge-on-Si heterosystem, $\mathrm{c\Gamma-LH}$ shall be the fundamental transition hence the low energy peak of the doublet.  
Relying on the spin-dependent selection rules for the dipole-allowed transitions at the zone center \cite{zutic:review2004} [see Fig~\ref{fig2}(a)], our findings suggests that the majority of the conduction carriers, involved in the radiative recombination, has spin up orientation. Moreover, since the higher energy feature of the doublet is absent in the 4K PL of n-Ge and it becomes more pronounced by increasing the acceptor concentration, we conclude that p-doping is effective in populating the energetically unfavored HH states. 

To put these findings on a more solid ground and to study further the origin of the exceptionally high $\rho$ observed in i-Ge, we evaluated the low temperature electronic band structure and the photogenerated spin population by means of a first-neighbor tight-binding Hamiltonian. To this aim we adopt the parametrization first proposed in Ref.~\onlinecite{Jancu:PRB07}, which includes $sp^3d^5s^*$ orbitals and spin-orbit interaction. The two main effects of strain, summarized by the calculated band structure shown in Fig.~\ref{fig2}(b) and (c), are the removal of the LH/HH degeneracy and the band gap shrinking. 
At the value of $\varepsilon_\parallel$=0.25\%, used in our calculations to model the thermal strain of Ge-on-Si at 4 K, the expected LH-HH splitting is 16 meV [see Fig.~\ref{fig2}(b)]. 
This is consistent with the separation of $\approx 19$ meV measured for the cross-polarized PL lines of i-Ge [see Fig.~\ref{fig1}(b)], and it underlines the potential of polarization-resolved PL spectroscopy to probe the strain-induced changes of the electronic structure. 

Moreover, Fig.~\ref{fig2}(c) elucidates that the main effect of the band gap narrowing on the optical orientation process is that the absorption of photons at 1.165 eV provides a sizable access to the $\mathrm{c\Gamma-SO}$ transition. The spin up orientation of electrons photoexcited  from SO states can have an unexpected role in determining $\rho$ associated to the radiative recombination across the direct gap. 
The rationale for this is that electrons which originate from the first two topmost  VB states (V1 and V2)\cite{note} are promoted within the $\Gamma$-valley out of the zone center [see Fig.~\ref{fig2}(c)]. These electrons will thus posses a much larger kinetic energy with respect to the ones optically coupled to the SO states. 
More precisely, at the adopted pump energy of 1.165 eV, the transitions involving V1 and V2 states produce CB electrons with typical excess energy, measured from the bottom edge of the $\Gamma$-valley, of about 180 and 280 meV, respectively. On the other hand, electrons promoted from the SO band have only a few meV of kinetic energy [see top panel of Fig.~\ref{fig2}(d)]. To evaluate the non-trivial energy dependence of the spin population in the $\Gamma$-valley, we plot in Fig.~\ref{fig2}(e), as function of the $k_\parallel$ and $k_\perp$ components of the electron momentum, the quantity $\Theta=\mathrm{\delta(\hbar\omega_{p}-E_{tr})\vert \vec{p}\cdot\hat e_{\sigma+}\vert^2<S_z> d^3}k$, which represents the $\mathrm{\vec{k}}$-resolved contribution to the non-equilibrium spin ensemble. Here $\parallel$ and $\perp$ are relative to the light propagation direction z, $\mathrm{E_{tr}}$ is the transition energy, $\mathrm{<S_z>}$ is the expectation value of the component of the spin operator along z evaluated for the CB states. Finally, $\mathrm{\vec{p}}$ is the momentum operator and $\mathrm{\hat e_{\sigma+}}$ is the polarization versor corresponding to the right-handed circular polarization.
In line with the above discussion, we find that SO electrons are characterized by a small momentum, while transitions involving V1 and V2, clearly distinguishable in Fig.~\ref{fig2}(e), possess larger $k$ values.
It is useful to remember that spin-dependent selection rules strictly hold at the $\Gamma$ point, where initial and final states have a well-defined $\mathrm{J_z}$ quantum number.\cite{zutic:review2004} In our case, the VB states involved in the optical orientation process feature relevant $\mathrm{J_z}$ mixing, due to both strain and excess pump energy. 
For this reason, the spin-dependent selection rules hold only for the small momentum SO states which in fact result to be coupled with spin up CB levels [see Fig.~\ref{fig2}(a)]. At the same time, in the regions of the $(k_\parallel,k_\perp)$ plane where the main orbital symmetry character of the V1 and V2 states is HH (LH), the corresponding spin of the electron results to be down (up).

Once clarified the role of the V1, V2 and SO bands, we plot in the bottom panel of Fig.~\ref{fig2}(d) the energy resolved electron spin polarization, $\mathrm{P_s = \frac{n_{\uparrow}-n_{\downarrow}}{n_{\uparrow}+n{\downarrow}}}$,  evaluated taking into account the spin up ($\mathrm{n_{\uparrow}}$)  and spin down ($\mathrm{n_{\downarrow}}$) electronic population up to a given excess energy in the $\Gamma$-valley. Calculations reported in Fig.~\ref{fig2}(d) indicate that transitions involving V1 and V2 states cause a lowering of the positive spin polarization generated by the SO transitions. A negative sign for $\mathrm{P_s}$ is predicted for an excess energy larger than 180 meV. Notice, as shown in Fig.~\ref{fig2}(d), that the fraction of carriers excited from the top of the SO band is only about 4\%, due to the much larger joint density of states associated to the transitions from the V1 and V2 states occurring out of the zone center. Indeed the overall polarization evaluated integrating over the total spin ensemble at the $\Gamma$-valley turns out to be $\mathrm{P_s^{tot}}$=-65\%. Nevertheless, due to the spin-dependent selection rules for the optical transitions [Fig.~\ref{fig2}(a)], the helicity of the $\mathrm{c\Gamma-LH}$ and $\mathrm{c\Gamma-HH}$ PL features unambiguously demonstrates that the spectra are dominated just by the small fraction of spin up electrons optically coupled to the SO states.

\begin{figure*}
\includegraphics[width=2\columnwidth]{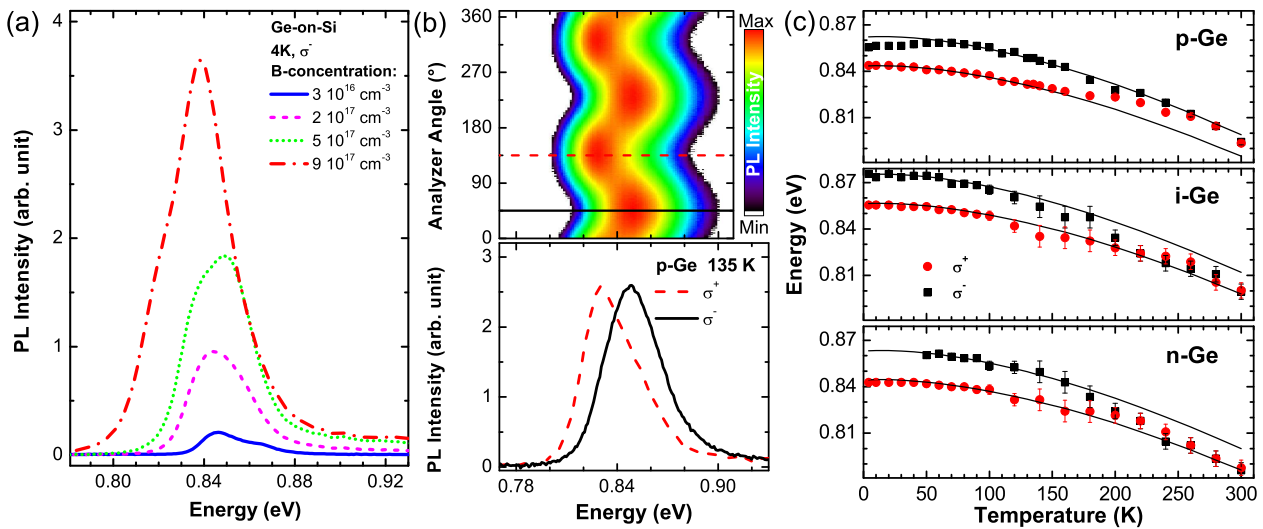}
\caption{\label{fig3} (a) $\sigma^{-}$ component of the 4 K PL spectra of p-type Ge-on-Si samples with different acceptor concentrations, namely $3\cdot 10^{16}$ (solid blue line), $2\cdot 10^{17}$ (dashed magenta line), $5\cdot 10^{17}$ (dotted green line) and $9\cdot 10^{17}$ $\mathrm{cm^{-3}}$ (dashed-dotted red line). (b) Color-coded map as a function of the analyzer angle (upper panel) and corresponding $\sigma^{-}$ (black solid line) and  $\sigma^{+}$ (red dashed line) PL spectra measured at 135 K in p-Ge (bottom panel). (c) Temperature dependence of the PL peak positions resolved for the $\sigma^{+}$ (red circled dot) and $\sigma^{-}$(black squared dot) components of the emission of p-Ge (upper panel), i-Ge (middle panel) and n-Ge (lower panel). Solid lines accounts for the expected $\mathrm{c\Gamma-LH}$ and $\mathrm{c\Gamma-HH}$ transition energies owing to the corresponding temperature \cite{varshni}, strain \cite{ishikawa:JAP2005} and doping \cite{jain:bgn1991} contributions.}
\end{figure*}

To gain a better understanding of this finding, we shall keep in mind that only radiative transitions across the direct gap with very small electron momentum are active, since at the adopted excitation density and doping levels, the quasi Fermi level in the VB of intrinsic and doped Ge is of the order of 10 meV, which corresponds to active transitions with conduction excess energy of less than about 100 meV. Above all, the steady-state spin population, which results from the relaxation of the photogenerated carriers due to both inter-valley scattering in the CB and the radiative and non radiative recombinations, may significantly differ from the calculated photogenerated distribution. Previous results reported for bulk unstrained Ge \cite{Mak94, Zhou94, pezzoliPRB2013, Pezzoli15} demonstrate that $\Gamma$-valley electrons with large kinetic energy are more effectively scattered out of the $\Gamma$-valley into lower energy X and L satellite valleys, finally suppressing their contribution to the radiative recombination events across the direct gap. These arguments for the Ge-on-Si heterostructures are remarkably corroborated by the striking agreement between the PL  polarization of +85\% measured for the fundamental transition in i-Ge and the predicted spin polarization of +90\% for the subset of electrons photogenerated from the SO band [Fig.~\ref{fig2}(d)]. 
In addition to the spin-selective dwell of electrons in the $\Gamma$-valley, it is worth noticing that biaxial tensile strain is ultimately beneficial for the achievement of high $\rho$ in the Ge-on-Si heterosystems. As opposed to the relaxed bulk case, strain lifts the LH/HH degeneracy leading to the spectral separation of photons emitted with opposite helicity via the otherwise interfering $\mathrm{c\Gamma-LH}$ and $\mathrm{c\Gamma-HH}$ transitions.
This matching between the measured $\rho$ and the calculated $\mathrm{P_s}$ along with the highly-mixed spin states expected for the photogenerated hole ensemble strongly suggest that holes participating to the radiative recombination events can be assumed to be unpolarized. Although no direct measurement is provided for the spin relaxation time of holes, it can be inferred, in light of the above discussion, that is should be shorter than the intervalley scattering time in the CB and thus in the hundreds-of-fs time range. \cite{Zhou94, loren:PRB2011}

Additional data supporting the physical picture leading to the PL polarization discussed above come from the analysis of the doped samples. Indeed, back-scattering of high energy electrons from the satellite X valleys towards the zone center can be strongly enhanced by Coulomb interaction with doping-induced charges.\cite{pezzoliPRB2013, Pezzoli15} This mechanism is known to result in doping-related changes of the PL polarization due to the increased accumulation at the edge of the $\Gamma$-valley of the spin down electrons originally photogenerated from the V1 and V2 states.\cite{pezzoliPRB2013, Pezzoli15} Consistently with this mechanism, we indeed measured $\rho$ of the fundamental transition that in n-Ge and p-Ge are smaller than in the less doped i-Ge.

We carried out a systematic polarization-resolved PL study as a function of doping and lattice temperature. Fig.~\ref{fig3}(a) shows the 4 K PL spectra of Ge-on-Si samples with an increasing acceptor concentration. As expected, there is a red-shift of the peak position due to doping-induced band gap narrowing and an overall enhancement of the PL intensity with the B content.\cite{jain:bgn1991, pezzoliPRB2013} However, the $\sigma^{-}$ component of the PL, shown in Fig.~\ref{fig3}(a), points out a rapid growth of the $\mathrm{c\Gamma-HH}$ transition, which becomes dominant above $4\times 10^{17}$ $\mathrm{cm^{-3}}$, and results in an overall peak circular polarization that is counter polarized with respect to the laser excitation. Indeed, the Fermi level shifts deeper in the VB by increasing the density of extrinsic holes, thus providing access to lower energy electron states and causing a rise of the population of the HH band, which is optically coupled to spin up electrons by emission of $\sigma^{-}$ photons. 

Temperature can play a similar role in modifying the occupation of the available hole states, eventually changing the spectral weight between $\mathrm{c\Gamma-LH}$ and $\mathrm{c\Gamma-HH}$, and favoring the latter in the high temperature regime even for less p-doped samples. Fig.~\ref{fig3}(b) shows the modulation pattern of the PL intensity measured in p-Ge at 135 K. The polarization analysis allows us to unfold the presence of the $\mathrm{c\Gamma-LH}$ and $\mathrm{c\Gamma-HH}$ lines, demonstrating their similar intensity, that is the equal weight of the $\sigma^{+}$ and $\sigma^{-}$ PL components. At this temperature the thermal energy, which corresponds to 12 meV, is quite close to the energy separation between LH and HH at the $\Gamma$-point. Besides thermal occupation, inward scattering of the photogenerated LH might be active as well. Above 135 K, the $\sigma^{-}$-polarized $\mathrm{c\Gamma-HH}$ contribution gains spectral weight over the $\sigma^{+}$-polarized $\mathrm{c\Gamma-LH}$, and when the temperature exceeds 200 K, these two components merge into a broad single line. Above this temperature, the PL peak shows a vanishing $\sigma^{-}$ polarization, that is counterpolarized with respect to the excitation. In the following, we compare the measured temperature dependence of the PL peak positions for the $\sigma^{+}$ and $\sigma^{-}$ polarization-resolved spectra and the predicted $\mathrm{c\Gamma-LH}$ and $\mathrm{c\Gamma-HH}$ transition energies. The latter ones have been obtained by considering the band-gap induced changes due to temperature \cite{varshni} and doping \cite{jain:bgn1991}, and the concomitant strain-induced red-shift and VB splitting \cite{ishikawa:JAP2005} at a given lattice temperature. The kT/2 factor accounting for the difference between the direct gap energy and the PL peak maxima has also been considered.\cite{pavesi94} The upper panel of Fig.~\ref{fig3}(c) shows the peak positions for the $\sigma^{+}$ (full dots) and $\sigma^{-}$ (full squares) PL components of p-Ge in the 4 to 300 K range. Below 200 K, the peak positions agree well with the predicted ones (solid line), while, above 200 K the $\sigma^{+}$ data collapse on the higher energy component of the doublet, thus corroborating in p-Ge the dominance of $\mathrm{c\Gamma-HH}$ in the high temperature regime.   

Remarkably, Fig.~\ref{fig3}(c) demonstrates that, as opposed to p-doped Ge-on-Si, $\mathrm{c\Gamma-LH}$ provides the main contribution to the high temperature PL of i-Ge and n-Ge. Even though above 200 K the PL of these two samples becomes not circularly polarized, the peak positions are fully consistent with the lower energy component of the doublet, namely the $\mathrm{c\Gamma-LH}$ transition. Such finding unravels the relative role of LH and HH occupancy in determining the PL spectral shape of tensile-strained Ge. This has been largely neglected in previous literature works \cite{elkurdi:APL2010, boztug:NN2014, harris:JAP2014},  although it has a crucial importance for the precise evaluation of the optical gain. \cite{boztug:NN2014} 

In conclusion, we elucidated the spin-dependent optical properties of a strained multivalley material, such Ge-on-Si, obtaining high PL polarizations degrees even for a large energy detuning of the laser pump. Noticeably, the simultaneous access to the $\mathrm{c\Gamma-LH}$ and $\mathrm{c\Gamma-HH}$ transitions allowed us to directly disclose their relative spectral weight and to put forward a spectroscopic approach for the measurement of strain and deformation potentials.

\begin{acknowledgments}
The authors would like to acknowledge N. Severino for technical assistance and M. Guzzi for fruitful discussions. This work was supported by the Fondazione Cariplo through the Grant $2013-0623$.
\end{acknowledgments}


\begin{thebibliography}{36}
\bibitem [{V1 and V2 symbols are adopted to identify the topmost and second
	valence states optically coupled to the conduction out of the zone center
	while HH, LH and SO are used for the $\Gamma$-point levels()}]{note}%
\BibitemOpen
V1 and V2 symbols are adopted to identify the topmost and second valence
states optically coupled to the conduction out of the zone center while HH,
LH and SO are used for the $\Gamma$-point levels,\ \href@noop {}
{}\BibitemShut {NoStop}%
\end{thebibliography}
%

\end{document}